\documentclass[12pt,preprint]{aastex}
\usepackage{natbib}
\usepackage{CJK}

\newcommand{\myemail}{guoyang@nju.edu.cn}
\newcommand{\jgrs}   {J. Geophys. Res. (Space Phys.)}

\begin{document}
\begin{CJK*}{UTF8}{gbsn}
\title{Evolution of Hard X-ray Sources and Ultraviolet Solar Flare Ribbons for a Confined Eruption of a Magnetic Flux Rope}
\author{Y. Guo (郭洋)$^{1,2}$, M. D. Ding (丁明德)$^{1,2}$, B. Schmieder$^{3}$, P. D\'emoulin$^{3}$, H. Li (黎辉)$^{4}$}

\affil{$^1$ Department of Astronomy, Nanjing University, Nanjing 210093, China} \email{\myemail}
\affil{$^2$ Key Laboratory for Modern Astronomy and Astrophysics (Nanjing University), Ministry of Education, Nanjing 210093, China}
\affil{$^3$ LESIA, Observatoire de Paris, CNRS, UPMC, Universit\'e Paris Diderot, 5 place Jules Janssen, 92190 Meudon, France}
\affil{$^4$ Purple Mountain Observatory, Chinese Academy of Sciences, Nanjing 210008, China}

\begin{abstract}
We study the magnetic field structures of hard X-ray sources and flare ribbons
of the M1.1 flare in active region NOAA 10767 on 2005 May 27.  We have found in 
a nonlinear force-free field extrapolation, over the same polarity inversion 
line, a small pre-eruptive magnetic flux rope located next to sheared magnetic
arcades. \textit{Ramaty High Energy Solar Spectroscopic Imager} (\textit{RHESSI}) 
and \textit{Transition Region and Coronal Explorer} (\textit{TRACE}) observed 
this confined flare in the X-ray bands and ultraviolet (UV) 1600 \AA \ bands, 
respectively. In this event magnetic reconnection occurred at several locations. 
It first started at the location of the pre-eruptive flux rope. Then, the observations 
indicate that magnetic reconnection occurred between the pre-eruptive magnetic 
flux rope and the sheared magnetic arcades more than 10 minutes before the flare peak.
It implied the formation of the larger flux rope, as observed with \textit{TRACE}. 
Next, hard X-ray (HXR) sources appeared at the footpoints of this larger flux 
rope at the peak of the flare. The associated high-energy particles may have been 
accelerated below the flux rope, in or around a reconnection region. Still, the
close spatial association between the HXR sources and the flux rope footpoints
favors an acceleration within the flux rope. Finally, a topological analysis of 
a large solar region including the active regions NOAA 10766 and 10767 shows the 
existence of large-scale Quasi-Separatrix Layers (QSLs) before the eruption of 
the flux rope. No enhanced emission was found at these QSLs during the flare, 
but the UV flare ribbons stopped at the border of the closest large-scale QSL.

\end{abstract}

\keywords{Sun: flares --- Sun: magnetic topology --- Sun: UV radiation --- Sun: X-rays, gamma rays}

\section{Introduction}
The process of a two ribbon flare is usually described by the CSHKP or the standard 
flare model \citep{1964Carmichael,1966Sturrock,1974Hirayama,1976Kopp}, which has 
been extended in various ways by many authors. The generally accepted view is summarized below. 
Before the occurrence of a flare, a core field with highly sheared field lines or a magnetic flux 
rope lies below the overlying arched envelope field. Due to the onset of a magnetic 
instability, the core field starts to rise and stretches the envelope field to 
form a current sheet below it. The magnetic reconnection in the current sheet 
converts the magnetic energy into kinetic and thermal energies of plasma and 
particles, which propagate along the reconnected field lines below the reconnection
site and generate soft X-ray loops along the magnetic arcades and hard X-ray (HXR) 
sources at the footpoints of the loops. The magnetic reconnection site moves upward as 
the reconnection proceeds, which generates new soft X-ray loop shells above the
older ones that have cooled down to extreme ultraviolet (EUV) and H$\alpha$ loops. 
The intersection of the loop system with the chromosphere and transition region 
displays the pattern of flare ribbons. The eruption of the magnetic flux rope above 
the reconnection site may propel plasma into the interplanetary space and form 
a coronal mass ejection (CME) if the eruption is not confined to the low corona (because
of a too strong overlying magnetic arcade).

However, there is one puzzling problem in the observations of flare ribbons
and HXR sources at the loop footpoints. While flare ribbons observed in ultraviolet 
(UV) and H$\alpha$ bands appear as elongated brightening structures on both sides of a 
polarity inversion line of the associated line-of-sight magnetic field, ribbon-like 
HXR sources have only been reported in very rare cases \citep{2001Masuda,2007LiuC,2007Jing}.
Most HXR sources appear as compact point-like sources. The problem of lacking 
ribbon-like HXR sources is explained by the fact that electrons are most efficiently
accelerated in particular loops due to fast reconnection rate; therefore, weak 
HXR emissions cannot be recorded by present HXR instruments with limited dynamic 
ranges \citep{2002Asai,2007Temmer,2007Miklenic}. But the reason is still not clear 
why the reconnection rate is faster in that particular site.

Magnetic flux ropes serve as a promising candidate to produce HXR sources, since they 
play an important role in models of solar active phenomena 
including flares, filaments/prominences, and CMEs. There is more and more evidence 
showing that magnetic reconnection could also occur in the leading edge of an erupting 
flux rope, in addition to the classical current sheet tracing behind and stretched by it, 
both from observations \citep{2003Ji,2009Wang,2011Huang} and from numerical simulations
\citep{2003Amari,2003Roussev,2005Torok}. Especially, \citet{2009Wang} found that 
EUV brightenings always appear at the two far footpoints of erupting filaments with the 
Extreme-ultraviolet Imaging Telescope (EIT; \citealt{1995Delaboudiniere}) on board the
\textit{Solar and Heliospheric Observatory} (\textit{SOHO}) in regions of the quiescent 
Sun. This finding motivates us to study erupting flux ropes in active regions and to check 
if any brightenings appear at their footpoints. Moreover, we need to study the 
relationship between HXR sources and the flare ribbons.

The HXR and UV emissions in a flare are generated by energetic particles via different
radiation mechanisms. The high energy particles are accelerated in a suitable environment
produced by magnetic reconnection, which occurs preferably at magnetic null points, 
separatrices, or at least Quasi-Separatrix Layers (QSLs). QSLs refer to thin irregular 
volumes where the mapping of field lines, e.g. to the photosphere, has a drastic change
for a given three dimensional magnetic field. QSLs divide the magnetic field into different 
domains. These domains, however, may be continuously connected at some locations. This is 
different from separatrices associated to magnetic null points, where different domains are 
totally topologically distinct. 

\citet{1996Demoulin} proposed a method to compute the locations of QSLs. 
Given a three dimensional magnetic field in a volume, one integrates a field line 
from $P(x,y,z)$ to both directions with a distance $s$ on each side. Taking two
points $(x',y',z')$ and $(x'',y'',z'')$ on both ends of the field line, a vector 
can be defined as $\mathbf{D}(x,y,z) = \{X_1,X_2,X_3\}=\{x''-x',y''-y',z''-z'\}$. 
The vector $\mathbf{D}(x,y,z)$ changes drastically in QSLs given a small displacement 
of point $P(x,y,z)$. Thus, if the norm $N$ is defined as
\begin{equation}
N(x,y,z,s) = \sqrt{\displaystyle\sum\limits_{i=1,3} \left[
\left(\frac{\partial X_i}{\partial x} \right)^2 +
\left(\frac{\partial X_i}{\partial y} \right)^2 +
\left(\frac{\partial X_i}{\partial z} \right)^2 \right]} ,
\label{eqn:norm1}
\end{equation}
QSLs are field lines with $N \gg 1$. In a practical numerical computation, we need 
to compute $N$ at each point in a volume, which is very time consuming. 
Therefore, \citet{1996Demoulin} suggested to compute a fixed number of points from 
a coarse grid to finer and finer grids. 

Equation~(\ref{eqn:norm1}) can be further simplified if one limits the positions 
of the two ends with more restrictions. For instance, if they are line-tied on the 
photosphere where $z'=z''=0$, the partial derivatives of $X_3$ to any coordinate 
equal zero, and the footpoints of a field line only depend on $x$ and $y$ (but not 
$z$). Then, the partial derivatives of $X_i$ $(i=1,2,3)$ to $z$ are zero and 
Equation~(\ref{eqn:norm1}) is reduced to
\begin{equation}
N_\pm \equiv N(x_\pm,y_\pm) = \sqrt{ \left(\frac{\partial
X_\mp}{\partial x_\pm} \right)^2 + \left(\frac{\partial
X_\mp}{\partial y_\pm} \right)^2 + \left(\frac{\partial
Y_\mp}{\partial x_\pm} \right)^2 + \left(\frac{\partial
Y_\mp}{\partial y_\pm} \right)^2 } , \label{eqn:norm2}
\end{equation}
where $\{X_\mp, Y_\mp\} = \{x_\mp-x_\pm, y_\mp-y_\pm\}$ \citep{1995Priest}. 
\citet{2002Titov} pointed out that $N_+$ does not always equal to $N_-$ for 
the same field line, which are computed at the footpoints of the same field 
line, $(x_+,y_+)$ and $(x_-,y_-)$, respectively. They proposed the squashing 
degree $Q$ as the measure of field line mapping, and
\begin{equation}
Q = \frac{N_+^2}{|B_{n+}/B_{n-}|} = \frac{N_-^2}{|B_{n-}/B_{n+}|},
\end{equation}
where $B_{n+}$ and $B_{n-}$ are the normal components of the magnetic field
at the two ends of a field line. The advantage to define the squashing 
degree $Q$ instead of the norm $N$ is that it is symmetric in computations at 
both footpoints of a field line. QSLs are then defined as those field lines 
with $Q \gg 2$.

The M1.1 flare in active region NOAA 10767 on 2005 May 27 is a good sample for us
to study the relationship between HXR sources and UV ribbons. \citet{2010Guo2} has 
found a magnetic flux rope and dipped magnetic arcades coexisting along the H$\alpha$ 
filament in the active region with the nonlinear force-free field model. 
The chirality of the filament barb is left bearing 
in the magnetic arcade section with negative magnetic helicity, which would induce 
a filament barb with right bearing in the flux rope section with the same magnetic 
helicity. \citet{2010Guo3} found that the eruption of the flux rope was confined
in the corona. With a detailed analysis on the twist number and decay index of 
the background magnetic field, \citet{2010Guo3} concluded that the eruption was 
triggered and initially driven by the kink instability, but the background magnetic
field did not decrease fast enough with height, thus prevented the occurrence of 
an ejective eruption with an CME. 

In this paper, we study the HXR and UV emissions of the M1.1 flare 
on 2005 May 27. Particularly, we try to find their temporal and spatial relationships
and to link these emissions with the three dimensional magnetic field structure,
i.e. the computed flux rope and QSLs. Observations and data analysis are described in Section 
\ref{sec:obser}. Results are presented in Section \ref{sec:resul}. We discuss our findings 
in Section \ref{sec:discu} and draw our conclusions in Section \ref{sec:concl}.

\section{Observations and Data Analysis} \label{sec:obser}
The M1.1 flare that occurred in active region NOAA 10767 on 2005 May 27 was observed 
uninterruptedly by the \textit{Transition Region and Coronal Explorer} 
(\textit{TRACE}; \citealt{1999Handy}) in the 1600~\AA \ band with a cadence of 
$\sim$30 s during the whole flare time. We calibrate the observed data
by subtraction of the detector dark current and normalization with the 
flat field and the exposure time. The final derived data are in units of
DN~s$^{-1}$~pixel$^{-1}$. The X-ray observations of the M1.1 flare were obtained
by the \textit{Ramaty High Energy Solar Spectroscopic Imager} (\textit{RHESSI}; 
\citealt{2002Lin}), which is a space-borne instrument that provides imaging 
spectroscopy observations both in X-rays and gamma rays from 3 keV to 17 MeV. 
Nine rotating collimators with two grids at both ends of each collimator convert the 
spatial image of the Sun into the temporal modulation of the photon counts, 
which are recorded by nine germanium detectors respectively with high energy 
resolution ($\lesssim$1 keV at 3 keV to $\sim$5 keV at 5 MeV). The detailed 
analysis of X-ray data is described in the following section.

\subsection{\textit{RHESSI} Imaging and Imaging Spectroscopy} \label{sec:imgsp}

\begin{figure}
\includegraphics[width=0.5\textwidth]{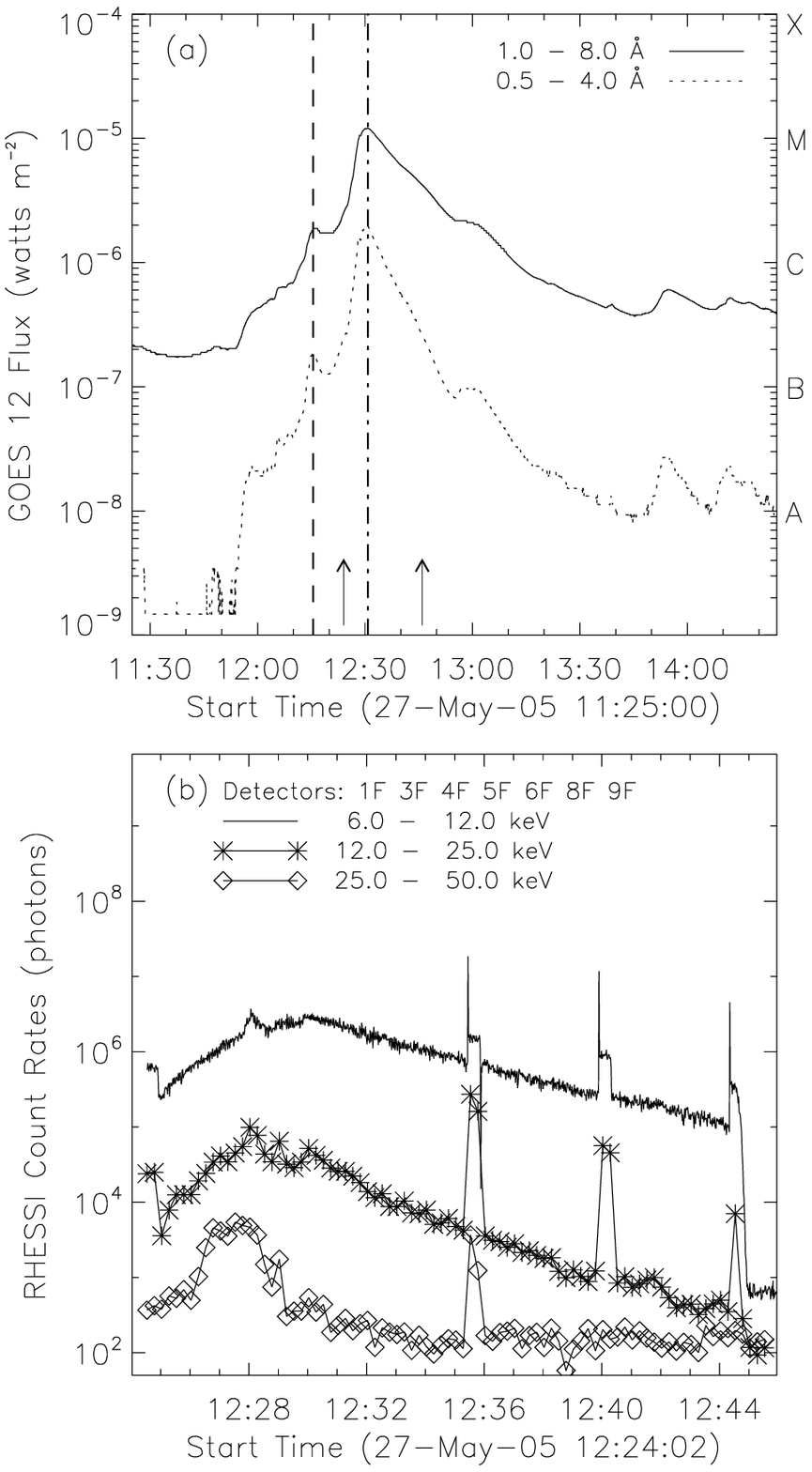}
\caption{(a) Soft X-ray flux from the \textit{GOES} 12 satellite of the  M1.1 flare 
on 2005 May 27. Dashed and dash-dotted lines indicate two peaks of the X-ray flux. 
The two arrows denote the time range of the \textit{RHESSI} light curve shown below.
(b) Corrected X-ray light curve obtained by \textit{RHESSI}. The peaks at different 
energy bands started at around 12:35, 12:40, and 12:44 UT are instrumental artifacts 
caused by removing the thicker attenuator before the detectors.} \label{fig:xflux}
\end{figure}

Figure \ref{fig:xflux}a displays the soft X-ray flux obtained by 
\textit{Geostationary Orbiting Environmental Satellites} (\textit{GOES}) 12, 
showing that the M1.1 flare peaked at $\sim$12:30 UT on 2005 May 27. The HXR count 
rates measured by \textit{RHESSI} at higher energy bands (i.e., 25.0--50.0 keV) 
reached their peaks at $\sim$12:28 UT as shown in Figure \ref{fig:xflux}b, 
about 2 minutes earlier than the peaks at the soft X-ray (SXR) bands. Such a 
time evolution behavior is due to the Neupert effect, i.e., the integral of 
the non-thermal fluxes coincides with the thermal fluxes. We have checked the 
time derivative of the \textit{GOES} flux at 1.0--8.0 \AA \ and the 
\textit{RHESSI} flux curve at 25.0--50.0 keV. Their peaks coincide with each 
other very well, which justifies the Neupert effect. We fit the spectra with
thermal and non-thermal components in fifteen 20-second accumulation intervals 
from 12:25:20--12:30:20 UT. It is found that the spectral indices in the non-thermal
component display a typical soft-hard-soft evolution.

\begin{figure}
\centering
\includegraphics[width=1.0\textwidth]{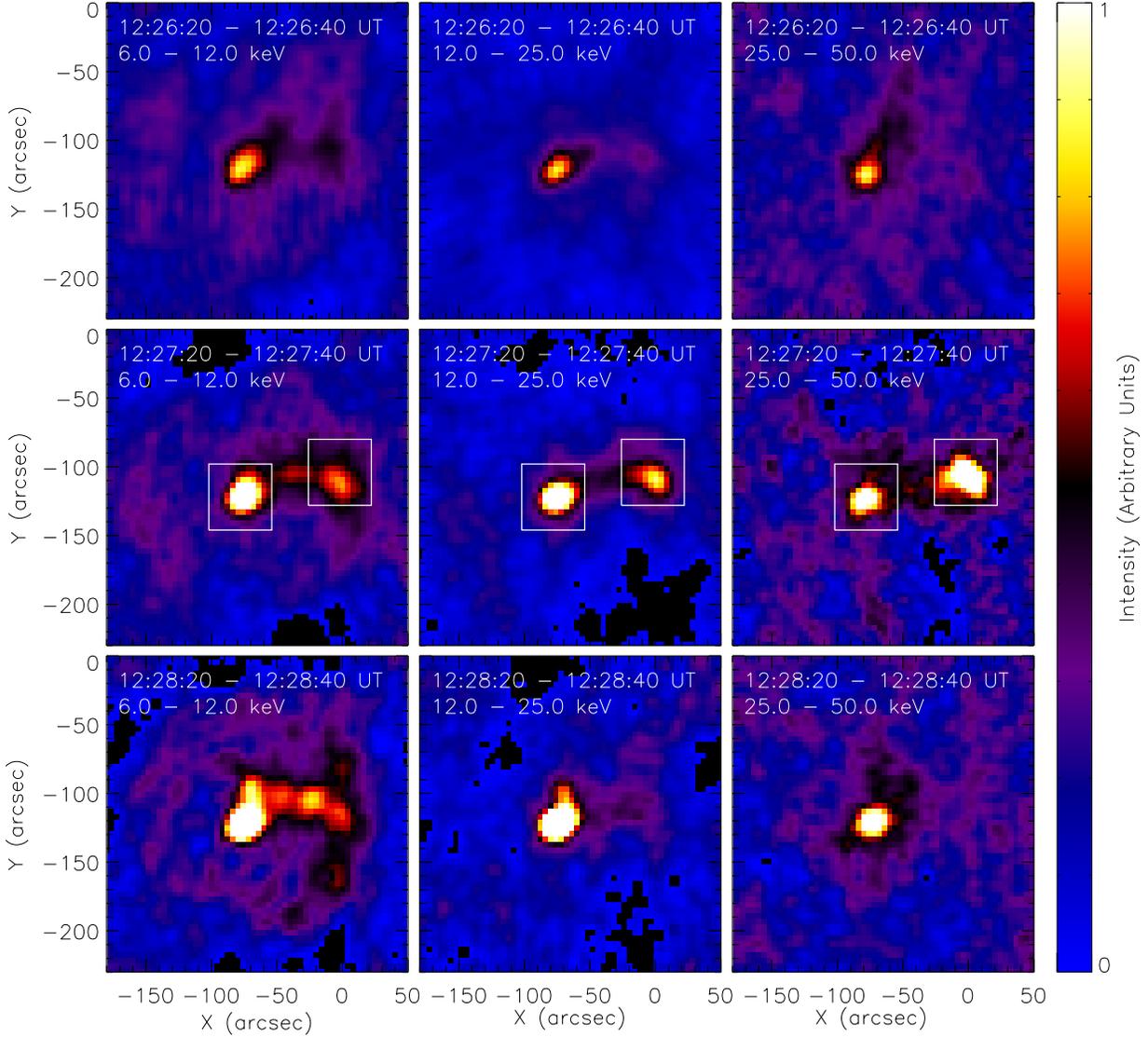}
\caption{X-ray images reconstructed from \textit{RHESSI} observations with the clean method 
in three energy bands (6.0--12.0, 12.0--25.0, and 25.0--50.0 keV) and three time intervals 
close to the peak of the M1.1 flare on 2005 May 27. The color-flux scale is the same 
in each column, but different within each row. Six detectors (3F--8F) are selected to 
reconstruct the images. The white boxes in the middle row enclose the regions in which 
the photon flux is integrated to build the spectra (Figure~\ref{fig:xspec}).} \label{fig:ximage}
\end{figure}

We plot the X-ray images reconstructed from \textit{RHESSI} observations with the clean 
method in three energy bands  (6.0--12.0, 12.0--25.0, and 25.0--50.0 keV) and three time 
intervals around the peak of the M1.1 flare in Figure~\ref{fig:ximage}. The 
figure shows that both footpoints appeared at the middle time. Only the eastern 
footpoint was present at all the three energy bands one minute before and after 
the middle time except the energy band of 6.0--12.0 keV at 12:28:20--12:28:40 U. 
In each of the three energy bands, the eastern and the 
western footpoints display different evolution behaviors in flux. The flux of 
the eastern footpoint increases monotonically in all the three energy bands 
with time, while the flux of the western footpoint increases monotonically 
only in the energy band of 6.0--12.0 keV (Figure~\ref{fig:ximage}). Indeed, the
flux of the western footpoint first increases and then decreases in the other
two energy bands. Finally, in the time interval of of 12:27:20--12:27:40 UT
and in the lower energy band (6.0--12.0 keV), the flux at the eastern footpoint
is larger than that at the western one, while the situation is reversed in the
higher energy band (25.0-50.0 keV, see the central row in Figure~\ref{fig:ximage}).

\begin{figure}
\centering
\includegraphics[width=0.8\textwidth]{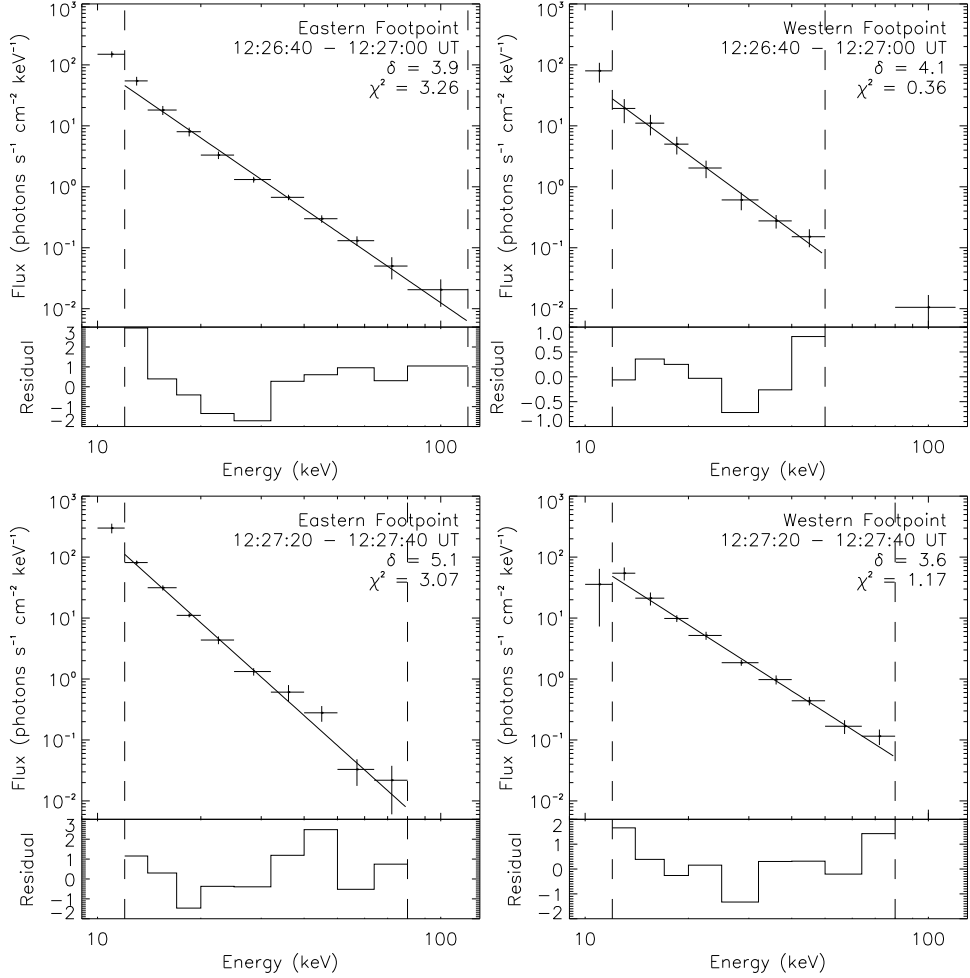}
\caption{Observed hard X-ray spectra with vertical and horizontal error bars showing 
the errors in the flux and the width of energy bins, respectively. The spectra
are constructed in the regions enclosed by the rectangular boxes as shown in 
Figure~\ref{fig:ximage}. They are fitted by a power law function (solid line), 
with the absolute value of the power index $\delta$ shown in each panel. Top and 
bottom rows show the spectra at two time intervals, i.e., 12:26:40--12:27:00 UT and 
12:27:20--12:27:40 UT, respectively. The two vertical dashed lines in each panel 
indicate the fitting energy ranges. The normalized residuals are shown at the bottom 
of each panel.} \label{fig:xspec}
\end{figure}

To give a quantitative measurement of the non-thermal spectra at each footpoint, 
the photon fluxes are integrated for sub-regions enclosed by the boxes as shown 
in the middle row of Figure~\ref{fig:ximage} for different time intervals and 
energy bands. We use 20-second accumulation intervals and 11 energy bands between 
10 and 120 keV. We find that the flux versus energy relationship can be well 
fitted by a power law for fifteen time intervals within 12:25:20--12:30:20 UT 
at the eastern footpoint; but at the western footpoint, it can only be fitted
for six time intervals within 12:26:40--12:28:40 UT. Figure~\ref{fig:xspec} shows 
the observed X-ray spectra and their fittings as an example. The non-thermal spectra 
for both the eastern and the western footpoints exhibit a soft-hard-soft evolution 
around the peak of the flare. If we assume that the photon flux has a power law 
form, $F(E) \sim E^{-\delta}$, the power law index $\delta$ reaches the smallest 
value of 3.9 in the time interval of 12:26:40--12:27:00 UT for the eastern 
footpoint, and of 3.6 in the time interval of 12:27:20--12:27:40 UT for the western 
footpoint. Therefore, the HXR spectra become the hardest at different times for 
different footpoints.

\subsection{Magnetic Field Extrapolations}

\begin{figure}
\centering
\includegraphics[width=1\textwidth]{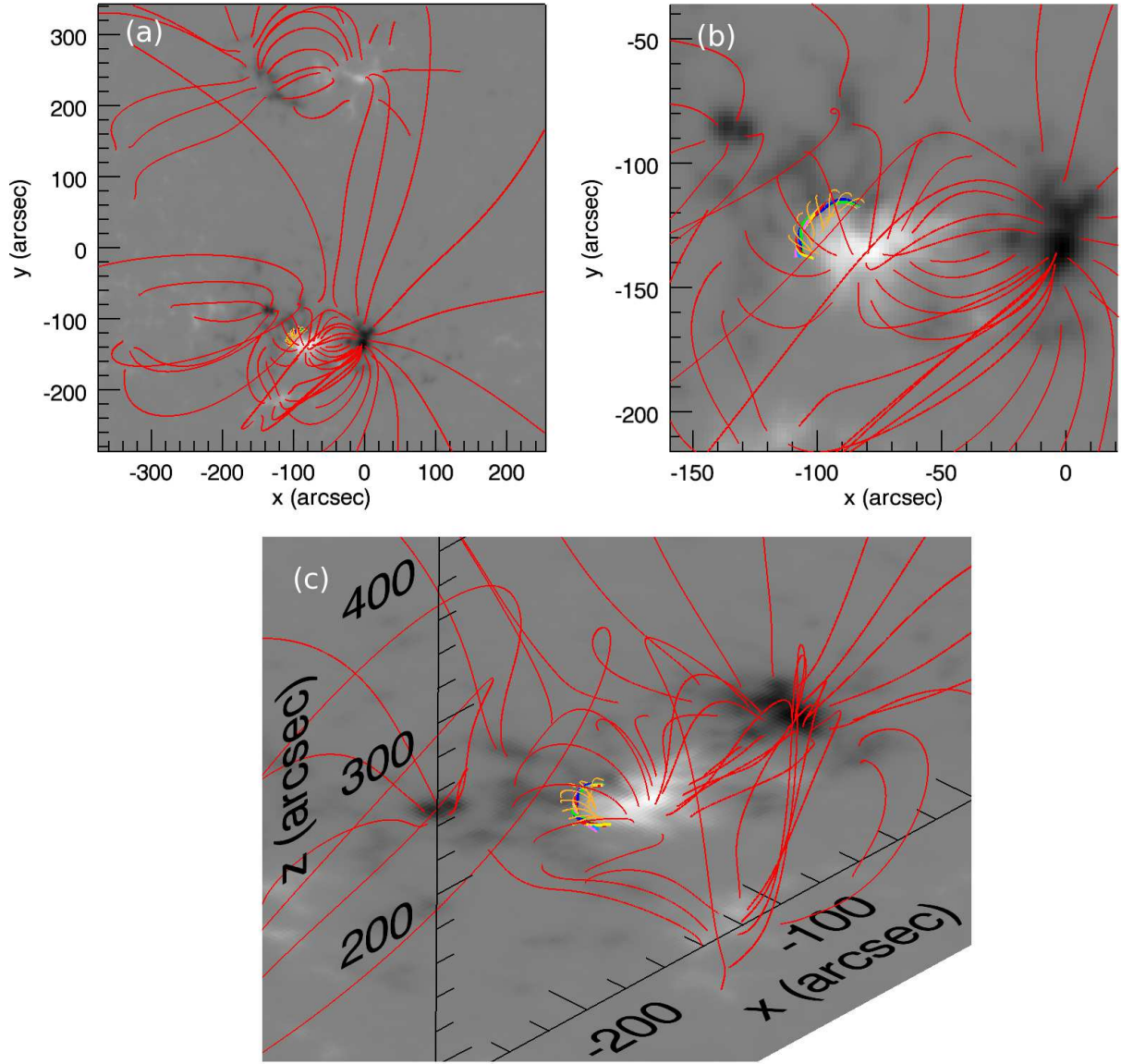}
\caption{Potential field extrapolation using MDI line-of-sight
magnetogram observed at 11:11 UT on 2005 May 27. The flux rope is
extrapolated by the nonlinear force-free field model with the
vector magnetic fields observed by \textit{THEMIS}/MTR at 10:17 UT on 2005
May 27. The field lines of the flux rope are overlaid with the
potential field after rotating the coordinates to the MDI
observation time. Different panels show different fields of views and viewing
angles.} \label{fig:potential}
\end{figure}

In order to compute QSLs, we construct a three dimensional magnetic field as 
shown in Figure~\ref{fig:potential} with the potential field model using the 
line-of-sight magnetic field observed by the Michelson Doppler Imager (MDI; 
\citealt{1995Scherrer}) on board \textit{SOHO}. The northern active region
NOAA 10766 and the southern one NOAA 10767 were observed at 11:11 UT on 2005 
May 27. \citet{2010Guo2} have analyzed active region NOAA 10767 by 
the nonlinear force-free field model. The bottom boundary is the vector 
magnetic field obtained by the \textit{T\'elescope H\'eliographique pour l'Etude 
du Magn\'etisme et des Instabilit\'es Solaires}/Multi-Raies (\textit{THEMIS}/MTR; 
\citealt{2007Bommier}). The field lines of the flux rope obtained by the nonlinear
force-free field model are overlaid with the potential field to show both the 
large and small scales of the magnetic field structure. Yet, we have not tried to 
construct a combined model incorporating both the nonlinear force-free field and 
the potential field here. Only selected field lines from both models are overlaid 
on the same figure for illustration. 

As shown in Figure~\ref{fig:trace}, the M1.1 flare occurred in the southern active 
region, which is connected to the northern one by the trans-equatorial field 
lines (Figure~\ref{fig:potential}a). In a thin layer at the border of this region 
with trans-equatorial connections, field lines have a drastic change of their 
linkages. We show in Figure~\ref{fig:qsl} that this thin layer is a QSL. Finally, 
from Figure~\ref{fig:potential}b and~\ref{fig:potential}c, we find that the region 
with highly twisted field lines is relatively small compared to the region covered 
with the surrounding arcade which extends up to the previous QSL.

\subsection{Coalignment of the Data}

We compare the \textit{RHESSI} observation with the \textit{TRACE} 1600~\AA \ 
image at the peak time of the M1.1 flare to check if the two HXR sources are 
conjugate footpoints of a flare. The two images have to be aligned with each other 
to compare their features. Because \textit{RHESSI} observes the full disk, the 
coordinates of the observed targets can be computed via the comparison of the 
positions of the solar limbs. The accuracy is within the spatial resolution of 
\textit{RHESSI} observations, which is $7''$ for images reconstructed from the 
six detectors 3F--8F. MDI also observes the full disk so that the coordinates 
of MDI observations can be precisely determined (with an error around its spatial
resolution of $2''$). 

The \textit{TRACE} image can be aligned with the magnetic field observed by 
\textit{THEMIS} by comparing the positions of the erupting feature in the 1600 
\AA \ band before the flare peak with the pre-eruptive flux rope found by the 
extrapolation (as shown in Figure \ref{fig:trace}a). Magnetic fields observed 
by \textit{THEMIS} and MDI are aligned by comparing their common features of 
the line-of-sight magnetic field. Thus, the \textit{TRACE} 1600 \AA \ image is 
aligned with the MDI observation. The alignment accuracy is estimated to be 
about $2''$, which is roughly the spatial resolution of MDI; by comparison, 
\textit{TRACE} and \textit{THEMIS} have much higher spatial resolutions of 
$0.5''$ and $0.8''$, respectively. The pointing offset of the \textit{TRACE} 
1600 \AA \ image can be used through the whole flare process after considering 
the solar rotation, since the pointing error was small during such a relatively 
short time range. Finally, the \textit{RHESSI} HXR image is aligned with the 
\textit{TRACE} 1600~\AA \ image as shown in Figure~\ref{fig:trace}c. Contours of 
the line-of-sight magnetic field, which were observed by \textit{THEMIS}/MTR at 
10:17 UT on 2005 May 27 and differentially rotated to 12:27 UT, are overlaid on 
both \textit{RHESSI} and \textit{TRACE} images. 

\section{Results} \label{sec:resul}

\subsection{Initial Presence and Development of a Flux Rope} \label{sec:devel}

\begin{figure}
\includegraphics[width=0.8\textwidth]{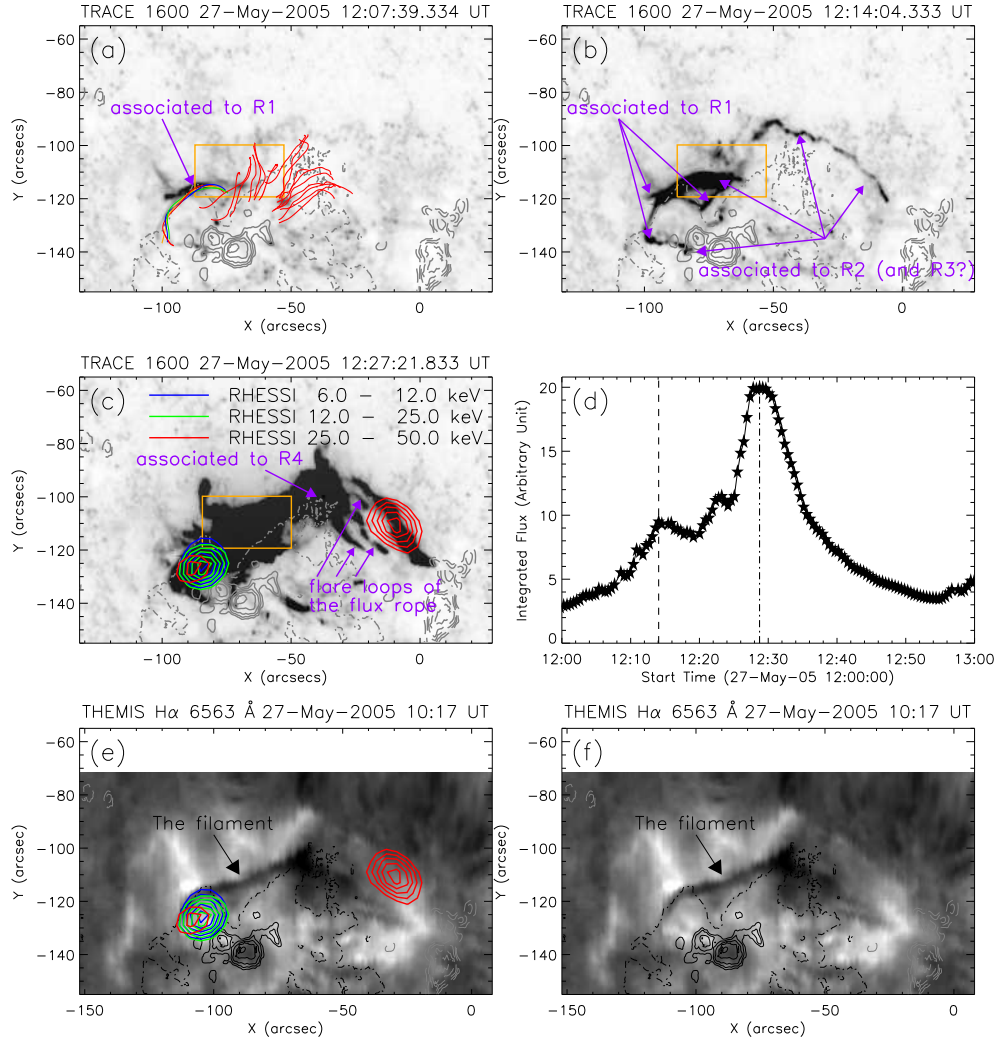}
\caption{(a) \textit{TRACE} 1600~\AA \ image (the gray scale is reversed) at 12:07 
UT overlaid by the pre-eruptive flux rope and some selected sheared field lines, 
which have been rotated to the observation time of the \textit{TRACE} image. 
Solid, dashed, and dash-dotted contours denote respectively the positive, 
negative polarities, and the polarity inversion line of the line-of-sight 
magnetic field observed by \textit{THEMIS}/MTR at 10:17 UT on 2005 May 27 
and rotated differentially to the observation time of the \textit{TRACE} image. 
The arrow points to a brightening region in the \textit{TRACE} 1600~\AA \ image. 
The labels R$i$ ($i=1$--4) are reconnection steps defined in Section~\ref{sec:steps}. 
(b) \textit{TRACE} 1600~\AA \ image at 12:14 UT. 
(c) \textit{TRACE} 1600~\AA \ image at the HXR peak time of the M1.1 flare overlaid 
by the \textit{RHESSI} X-ray contours. The integration time interval for the X-ray 
image is 12:27:20--12:27:40 UT. 
(d) \textit{TRACE} 1600~\AA \ flux integrated in the rectangular box as shown in previous 
panels. Dashed and dash-dotted lines indicate the two peaks of the integrated flux. 
(e) An H$\alpha$ filament overlaid by the \textit{RHESSI} X-ray contours as that
in panel (c) and being rotated to the observation time of the H$\alpha$ filament.
(f) The H$\alpha$ filament observed by \textit{THEMIS}/MTR on 2005 May 27.
} \label{fig:trace}
\flushleft
(An mpeg animation is available in the electronic edition of the journal.)
\end{figure}

\citet{2010Guo2} found a small flux rope in this active region about 2 hours 
before the peak of the flare by the nonlinear force-free field extrapolation 
method \citep{2000Wheatland,2004Wiegelmann}. The magnetic flux rope corresponded 
to the eastern part of an H$\alpha$ filament, whose western part was found by the 
magnetic extrapolation to be supported by sheared magnetic arcades. From the 
beginning of the flare, a first type of reconnection (called R1 hereafter) 
occurred nearby the flux rope. The consequence of such reconnection is the 
development of a bright feature along the polarity inversion line (see Figure 
\ref{fig:trace}a and the attached movie). As shown in Figure \ref{fig:trace}a, 
we find that the UV brightening generated by magnetic reconnection appeared 
nearby the pre-eruptive flux rope.

During the peak time of the confined eruption, the erupting magnetic flux rope, 
as suggested by the \textit{TRACE} 1600~\AA \ observation, was much longer than 
that at the eruption onset, as found by the nonlinear force-free field extrapolation. 
Magnetic reconnection (called R2 hereafter) might occur between the magnetic flux 
rope and the sheared arcades to form a longer flux rope, which facilitated the 
further eruption and the reconnection in the main phase. However, the above scenario 
should be taken with caution, since different nonlinear force-free field algorithms 
could not obtain a unique solution with observational data as the bottom boundary 
\citep[e.g.,][]{2009DeRosa}. Thus, we need to compare an extrapolated magnetic 
field with more observations, such as H$\alpha$ filaments and/or \textit{TRACE} 
171 \AA \ loops. \citet{2010Guo2} showed that the magnetic dips in the nonlinear 
force-free field model coincided with the locations of the associated H$\alpha$ 
filament, which is a test of the extrapolation result.

In order to find the evidence of the reconnection R2, it is better to check the 
HXR image evolution. Unfortunately, there were no \textit{RHESSI} data during 
11:54--12:25 UT and there was no clear evidence showing the onset reconnection 
at the center of the region in other time intervals. A bump on the \textit{GOES}
flux curve at 12:15 UT (Figure \ref{fig:xflux}a) shows an indirect evidence of 
energy release; however, we do not know where it comes from. Only the 
\textit{TRACE} 1600~\AA \ observation covered this active region during 
the flare. From these observations we find that there was no brightening 
at the location where the flux rope contacted with the sheared arcades 
before 12:00 UT. Next, we integrate the \textit{TRACE} 1600 \AA \ flux 
in the region surrounded by a box as plotted in Figure~\ref{fig:trace}a. 
The selected box tracks the region where the magnetic flux rope and the sheared
arcades contacted with each other and follows the solar rotation. The integrated 
flux curve is in the time range of 12:00--13:00 UT. As shown in Figure~\ref{fig:trace}d, 
a small peak appeared at 12:14 UT before the main peak of the integrated 
flux. Recall that at almost the same time, \textit{GOES} recorded a small bump in 
the SXR flux. The coincidence of the peak time suggests that the SXR
emission was generated in the same region as that of the 1600~\AA \ band, implying 
further that magnetic reconnection R2 possibly occurred at the center of the region at 
$\sim$12:14 UT (refer to Figure \ref{fig:trace}b) to form the finally erupted longer 
flux rope. The magnetic reconnection between the western magnetic arcades (tether 
cutting reconnection to build a larger flux rope, called R3 hereafter) may be 
initiated immediately after reconnection R2. There is a possible time overlap
when both R2 and R3 were in progress. 

We have presented a series of figures and a movie of the \textit{TRACE} 
1600~\AA \ observation in \citet{2010Guo3} to show the evolution of the helical 
rope-like structure. A movie showing the evolutions of \textit{TRACE} 1600~\AA \ 
images and \textit{RHESSI} X-ray sources is included in the electronic edition of the 
journal (the flux rope in Figure \ref{fig:trace}a, the arrows with labels in Figures 
\ref{fig:trace}a, \ref{fig:trace}b, and \ref{fig:trace}c, and 
Figure \ref{fig:trace}e are not shown in the movie). 
Three snapshots are displayed in Figures \ref{fig:trace}a, \ref{fig:trace}b, 
and \ref{fig:trace}c, respectively. It is found that the flare initiated 
at the location of pre-eruptive magnetic flux rope (Figure \ref{fig:trace}a).
The small pre-eruptive flux rope grew into a larger one after the reconnection R2 
(Figure \ref{fig:trace}b). Then, the erupting rope-like structure displayed an ascending 
and helically deforming evolution, which is a characteristic of the kink instability 
(Figure \ref{fig:trace}c), thus indicating the existence of a magnetic flux rope throughout 
the process of the flare. The rope-like structure stopped to ascend at a certain 
height and seemed not to evolve to any CME, suggesting that the eruption was confined 
in the corona.

\subsection{HXR Sources and the Magnetic Flux Rope}

Figure~\ref{fig:trace}c shows that the two \textit{RHESSI} HXR footpoints at 
25.0--50.0 keV are located at the two ends of a helical rope-like structure 
connecting them. We have concluded that the helical structure is an erupting 
flux rope that was formed by magnetic reconnection R2. Therefore, the HXR sources 
appeared at the footpoints of the erupting magnetic flux rope. Figure~\ref{fig:trace}e
shows the HXR sources overlaid on an H$\alpha$ image that was observed about 
two hours before the flare, which indicates that the western footpoint of the 
erupted flux rope is more extended to the west than the western footpoint of the
H$\alpha$ filament. This shift is coherent with the western extension of the flux
rope during the flare. Both reconnections R2 and R3 contributed to extend the flux 
rope to locations where no magnetic dips, so no filament, were present before the
flare. Two types of magnetic reconnection could be responsible for the generation 
of these HXR sources: reconnection R3 behind or reconnection (called R4 hereafter) 
within the erupting flux rope (Figure~\ref{fig:trace}c).

HXR sources at 25.0--50.0 keV first appeared at the eastern footpoint, and the energy 
spectra at the two footpoints are different from each other (Section~\ref{sec:imgsp}). 
The time when the two HXR sources became the hardest (at $\sim$12:27 UT) was earlier 
than the HXR flux peak time (at $\sim$12:28 UT), the integrated UV flux peak time 
(at $\sim$12:28 UT), and the SXR flux peak time (at $\sim$12:30 UT). Different HXR 
fluxes and spectral indices in the conjugate footpoints, which is usually termed as 
asymmetric footpoints, arise from the different properties in the process of particle 
acceleration and particle transport \citep[e.g.,][]{2009LiuW}. Interpretations of 
conjugate HXR footpoints require a detailed study of the particle acceleration 
mechanism, magnetic mirroring effect, column density of the flare loops and other 
effects, which is out of the scope of this paper.

\subsection{Flare Ribbons and Quasi-Separatrix Layers}

\begin{figure}
\centering
\includegraphics[width=1\textwidth]{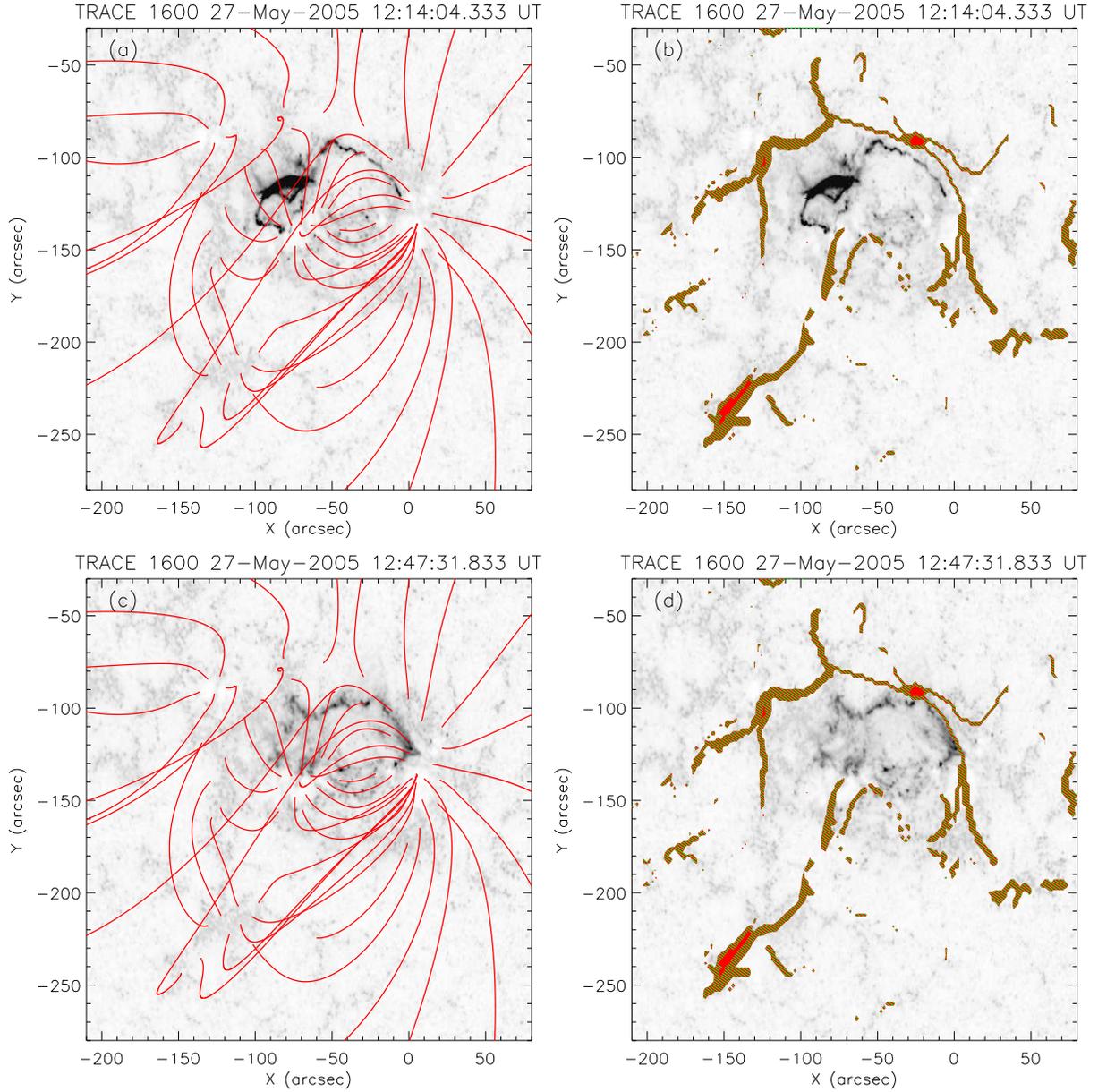}
\caption{(a) \textit{TRACE} 1600~\AA \ image at 12:14 UT overlaid with potential 
field lines at 11:11 UT, which have been rotated to the \textit{TRACE} observation 
time. (b) \textit{TRACE} 1600~\AA \ image at 12:14 UT overlaid with the intersection 
of QSLs with the photosphere. The QSLs are calculated with the potential field extrapolated 
with the MDI line-of-sight magnetogram at 11:11 UT. Only the QSL intersections with 
$Q \ge 10^4$ are shown as the hatched area. (c) \textit{TRACE} 1600~\AA \ image at 
12:47 UT overlaid with potential field lines at 11:11 UT, which have been rotated to the 
\textit{TRACE} observation time. (d) \textit{TRACE} 1600~\AA \ image at 
12:47 UT overlaid with the intersection of QSLs with the photosphere. The QSLs are 
the same to that in panel (b).} \label{fig:qsl}
\end{figure}

A \textit{TRACE} 1600 \AA \ image at 12:14 UT, when the flare ribbons clearly 
appeared, is overlaid with the potential field lines that are rotated to the 
\textit{TRACE} observation time as shown in Figure~\ref{fig:qsl}a. The 
intersection of computed QSLs and the photosphere is overlaid on the 
\textit{TRACE} 1600 \AA \ image in Figure~\ref{fig:qsl}b. The selected TRACE 
observation time is close to the first peak of the UV flux (Figure \ref{fig:trace}d) 
in the central area, so it was taken when reconnection R2 was occurring. The
northwestern ribbon has a distance of more than $5''$ to the location of the QSL 
intersection on the photosphere, which indicates that no field lines in the 
QSL computed with the potential field extrapolation were involved in the magnetic 
reconnection to produce the flare ribbons, in contrast with many previous studies which 
related flare ribbons to QSLs (see, e.g., \citealt{2007Demoulin}, and references
therein). In fact, the QSLs involved in R1 (and later in R2 and R3) are associated
to the presence of the erupting flux rope, which are not present in the potential field
extrapolation of Figure \ref{fig:qsl}. The potential-field QSLs are privilege 
locations of the concentrated current layers if there is enough magnetic field 
evolution in these regions \citep{2005Aulanier}. The absence of significant 
brightenings at the potential-field QSLs implies that the driving force by the 
distant erupting flux rope was not enough to build thin enough current layers 
(to be able to provide significant reconnection, so significant energy release).

As the flare proceeded, the flare ribbons separated apart. We overlay the potential
field lines and the intersection of the potential-field QSLs on the \textit{TRACE} 
1600 \AA \ image at 12:47 UT, the late phase of the flare (Figures \ref{fig:qsl}c 
and \ref{fig:qsl}d). The right part of the northwestern ribbon stopped at the border 
of the QSL, while the left stopped before reaching it. We interpret this result as 
follows. If a magnetic flux rope is ejected into the interplanetary space, the 
overlying arcade is fully stretched. Then, this arcade is expected to be fully 
reconnected and further build-up the ejected flux rope. In such a case, reconnection
is expected to transform all the arcade magnetic flux to the flux rope, then the
flare ribbons separate up to the arcade extension. However, if the event is confined, 
it is expected that a part of the overlying arcade stays (its downward magnetic 
tension confines the flux rope). Then, in this latter case, the flare ribbons are
expected to stop their progression before reaching the border of the arcade, so the
associated QSL. 

\subsection{Summary of the Reconnection Steps} \label{sec:steps}

We identify four steps of magnetic reconnection in the process of the flux 
rope eruption. The first reconnection, R1, occurred nearby the pre-eruptive
magnetic flux rope. Since the flare was observed from above and no height 
information can be obtained, we cannot determine whether R1 occurred below, 
within, or above the flux rope from the present observation. There are some
possible reconnection mechanisms for R1. First, It fits the general picture 
of the tether-cutting model with the progressive transformation of sheared 
arcades to a flux rope. This process is related to the work of \citet{2011Green}, 
where they presented a detailed study on the flux rope formation and eruption 
through flux cancellation in another active region. They showed that a flux 
rope can be formed by magnetic reconnection before its eruption. Secondly,
magnetic reconnection R1 may occur within the erupting flux rope. Or finally, 
the real case can also be the combination of the above two possibilities.

The second reconnection, R2, happened between the small flux rope in the eastern
part of the active region and the sheared magnetic arcades in the western part
(Figure \ref{fig:cartoon}a). This step started more than 10 minutes before the 
main UV and X-ray peaks of the flare. It built up a longer flux rope and probably 
stimulated the reconnection within the western sheared arcades (R3). It implied 
a western propagation of the flare brightenings along the polarity inversion
line. Such propagation was previously reported in another event \citep{2007Goff}, 
and their conclusion (e.g., their Figure 9) is similar to the one presented above.

The fourth step reconnection, R4, is pointed out by the presence of the HXR 
emission only at the footpoints of the erupting flux rope. While the spatial 
resolution and the intensity saturation of the observations do not permit to 
exclude that these HXR sources can be formed by reconnection at the periphery 
of the growing flux rope (so by reconnection R3), the UV observations point
to a strong energy release within the erupting flux rope.

Finally, the confined eruption of the flux rope did not have a large enough
effect on the large-scale QSLs (computed with a potential field) to build thin 
enough current layers, so to induce significant reconnection in the potential-field
QSLs (called R5 hereafter), as there was no significant brightenings associated 
to these QSLs.

\section{Discussion} \label{sec:discu}
\begin{figure}
\includegraphics[width=0.5\textwidth]{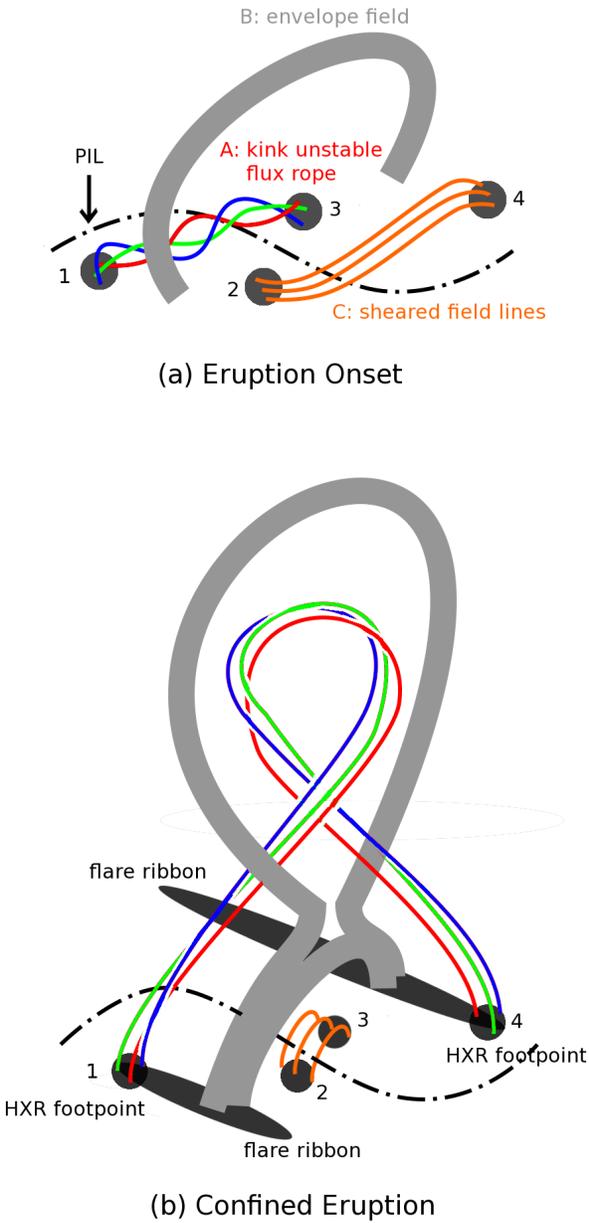}
\caption{Schematic picture of the onset and final stage of the confined eruption.
The idea is based on the tether-cutting model of \citet{2001Moore}. This picture
can be compared with the one for the ejective eruption in \citet{2007LiuC2}.} \label{fig:cartoon}
\end{figure}

We summarize the full scenario with a schematic picture in Figure~\ref{fig:cartoon}, 
where both the onset eruption and the final confined eruption are depicted. At the 
onset of the eruption as shown in Figure~\ref{fig:cartoon}a, the magnetic flux 
rope found by the nonlinear force-free field model built a highly non-potential 
state with a twist number exceeding the one suitable for the helical kink 
instability, which has been quantitatively analyzed in \citet{2010Guo3}. The 
helical kink instability is suggested to trigger and drive the eruption. In 
\citet{2010Guo3}, the authors also found that the final eruption was a confined 
one, i.e., the eruption of the magnetic flux rope was constrained within the
corona due to the large restoring force of the overlying magnetic field as shown 
in Figure~\ref{fig:cartoon}b. In this paper, we have mainly three findings.
First, the HXR sources appeared at the footpoints of a flux rope (as in the events
studied by \citet{2009LiuR} and \citet{2010Xu}). Secondly, the magnetic reconnection 
R2 occurred more than 10 minutes before the main peak of the flare. And thirdly, 
the UV flare ribbons stopped at the border of the potential-field QSLs. These 
findings have great implications on the mechanism and process of the M1.1 flare on 
2005 May 27, as discussed below.

First, we find that the conjugate HXR footpoints at the peak time of the 
flare were located at the two footpoints of a magnetic flux rope. This 
finding is different from the usual viewpoint, in particular what is based 
on the two dimensional flare models, where HXR footpoints are always located
at the footpoints of magnetic arcades below the reconnection site. These 
arcades are formed by magnetic reconnection of the envelope field, which is 
stretched by the erupting flux rope. Such post flare magnetic arcades are mostly
potential and perpendicular to the polarity inversion line, and so cannot provide 
a magnetic connection between the two HXR footpoints (Figure \ref{fig:trace}c and 
\ref{fig:cartoon}b). Such connection can only be provided by the flux rope 
connectivity (as deduced from \textit{TRACE} observations). It implies that high energy 
particles are preferably accelerated along the magnetic flux rope. The above 
results are related to the results of \citet{2011Cheng}. They found in another 
flare a very hot ($\sim$11 MK) ejected flux rope, which also suggests that a 
fast and effective heating mechanism is working.

Does magnetic reconnection only occur at the border of the flux rope or could it 
occur in the flux rope body? For the border reconnection case, tether cutting 
magnetic reconnection progressively transforms the surrounding arcade field 
lines to flux rope ones \citep[e.g.,][]{2004Torok}. So high energy particles 
and heating are input on all these newly formed field lines. If the initial flux
rope has a small extension compare to the one built up during the eruption, 
then most of the erupting flux rope would be filled with hot plasma and high
energy particles. In this process, most energy is provided only at the periphery
of the flux rope at a given time (some energy will be further provided by the
latter relaxation of the magnetic field). Alternately, reconnection within the 
flux rope could happen with an internal kink instability \citet{1997Galsgaard,
2007Haynes}. It requires that the twist is large enough within the flux rope so 
that the core becomes kink unstable. So far, this internal kink instability
has been proposed only for the heating of coronal loops since the instability 
does not affect much the external field (see above references). In the eruption
of the flare on 2005 May 27, an external kink instability is plausibly the cause
of the flux rope writhing as observed by \textit{TRACE} \citep{2010Guo3}. We
further propose here that an internal kink instability could drive internal 
reconnection which accelerate high energy particles. In this case, the HXR 
sources could be present within the flux rope footpoints, while in the case 
of tether cutting reconnection they should appear at the border of the footpoints. 
Due to the limitation of the spatial resolution of both UV and HXR observations 
in this study, these two cases cannot be discriminated.

Next, it is worthwhile to compare our findings with other studies on HXR sources and
UV ribbons in flares. \citet{2009LiuR} studied the HXR emissions in kinking filaments
for three cases on 2002 May 27, 2003 June 12, and 2004 November 10, respectively.
They found that there are two phases of eruptions, where compact HXR sources appear. 
In the first phase the sources appear at the endpoints of the associated filament, and 
in the second phase elongated ribbons appear at the footpoints of the magnetic arcades. 
The authors proposed that magnetic reconnection occurs between the two writhing filament 
legs, and later between the two envelope field legs (in the vertical current sheet) 
in the two phases, respectively. Our results are different from \citet{2009LiuR}
in two points. First, reconnection R1, R2, and R3 lead to the formation
of a larger flux rope that caused a confined eruption later; while in the events of
\citet{2009LiuR}, both phases of reconnection occurred at the time when the 
flux ropes have fully developed and writhed. Secondly, we find that the HXR sources
coincided with the footpoints of the flux rope at the HXR peak time. These sources 
did not move to the footpoints of magnetic arcades formed by magnetic reconnection
in the vertical current sheet as they were observed in the events on 2002 May 27 and 
2004 November 10 as shown in \citet{2009LiuR}.

Recently, \citet{2010Xu} found four HXR sources with \textit{RHESSI} at the onset 
stage of an X10 flare on 2003 October 29. The four sources are two conjugate pairs 
similar to the ones shown in Figure~\ref{fig:cartoon}a. This study provides additional 
evidence for the onset stage with two steps of magnetic reconnection. However, in our 
case, there was no observation with \textit{RHESSI} at the onset time of the M1.1 
flare. The difference between the two studies lies mainly in the behavior of the 
HXR sources. In the events of \citet{2010Xu}, the two outer sources (Sources 1 and 4 
in Figure \ref{fig:cartoon}a) disappeared at the peak time; while in our event, the 
two outer sources are the strongest and the two inner sources (Sources 2 and 3) were 
absent at the peak time. This difference implies that HXR sources could be formed 
both at the footpoints of the flare loops and/or of the erupted flux rope in 
different environments.

Finally, the northwestern ribbon in the UV 1600 \AA \ band appeared at a location 
with a detectable distance to the location of the intersection of the potential-field
QSL, and it stopped nearby the footpoints of the QSL. The intersection of the potential-field 
QSL on the photosphere was relatively stable during the impulsive energy release
process, since the shape of flare ribbons during the flaring time still mimic the 
shape of the QSL intersection on the photosphere that was observed about one hour 
before the flare. This is linked to the confined nature of this eruption, with a 
flux rope that did not succeed to overcome the downward magnetic tension of its 
overlying magnetic arcades. 

The UV flare ribbons were produced by magnetic reconnection R1, R2, and R3, which
is expected to occur in newly formed current layers during the eruption of the flux
rope (Figure \ref{fig:cartoon}b). As pointed out by \citet{2011Chen}, the QSLs 
associated to the above current layers are difficult to find with the present 
magnetic field extrapolation method (because it represents, at best, only the 
initial configuration). Finally, the erupting field was pushed close to the
large-scale potential-field QSL as the magnetic reconnection proceeded.

\section{Conclusions} \label{sec:concl}
We study the magnetic field structures of hard X-ray sources and flare ribbons of 
the M1.1 flare in active region NOAA 10767 on 2005 May 27. \citet{2010Guo2} has 
found a small pre-eruptive magnetic flux rope coexisting with sheared magnetic 
arcades in a nonlinear force-free field extrapolation. The observations indicate 
that this flare involved a multi-reconnection sites, as follows. First, 
\textit{TRACE} 1600~\AA \ and \textit{GOES} SXR fluxes suggest that an onset
magnetic reconnection occurred nearby the flux rope. This reconnection was triggered
and driven by the activation of the pre-eruptive magnetic flux rope, and it further
facilitated the flux rope eruption. Secondly, later on reconnection occurred between 
the pre-eruptive magnetic flux rope and sheared magnetic arcades more than 10 minutes 
before the flare peak time. Magnetic reconnection steps R2 and R3 provide a possible 
explanation for the formation of the larger flux rope observed by \textit{TRACE}. 
But we cannot exclude other possibilities due to the limitation of the data 
available and the nonlinear force-free field extrapolation. On one hand, there 
were no HXR observations at the early phase of the eruption, neither were there any EUV 
and SXR observations at this phase. On the other hand, the magnetic filed configuration 
obtained from the nonlinear force-free field should be taken with caution as we have 
discussed in Section~\ref{sec:devel}.

\textit{RHESSI} and \textit{TRACE} observations show that HXR sources appeared 
at the footpoints of the larger flux rope at the peak of the flare. We could not 
determine whether these sources were created by particles accelerated within or 
nearby the border of the large flux rope. Still, the spatial coincidence between 
the HXR sources and the footpoints of the flux rope favors particle acceleration 
within the flux rope. A possible mechanism could be the development of an internal 
kink instability, since it would induce the formation of a thin current layer, 
then of reconnection, within the flux rope.

Finally, a topological analysis of a large solar region including the active 
regions NOAA 10766 and 10767 shows the existence of large-scale QSLs before the 
eruption of the flux rope. Such QSLs did not participate in the flare, but the
extension of the flare ribbons is found to be confined inside the closest large-scale 
QSL computed from a potential field extrapolation. We conclude that the reconnection, 
involved in the confined eruption of the flux rope, was not involving larger scale 
structures than the arcade overlying the flux rope. The northwestern ribbon ended 
along the closest QSL computed with the potential field from a magnetogram taken 
before the flare. Such spatial coincidence indicates that the magnetic field should 
not deviate much from the potential field in the envelope field far from the core 
field region. The nonlinear force-free field model from the optimization method as 
derived in \citet{2010Guo2} has a smaller spatial extention than the above potential field 
extrapolation because of the limited field of view of the vector magnetogram available.  
Still, the nonlinear model indicates that the magnetic field gradually gets closer 
to the potential field as the distance from the center of the active region increases.  
Together with the good correspondence found previously between the extensions of the 
computed magnetic dips and the H${\alpha}$ filament, this is a confirmation that 
the nonlinear force-free field model provides a reliable approximation of the coronal field.

\acknowledgments
The authors thank the referee for helpful comments that improved the clarity of
the paper. Y.G. thanks Pengfei Chen very much for useful discussions. We are grateful 
to the {\it GOES}, {\it RHESSI}, {\it SOHO}, {\it THEMIS}, and {\it TRACE}
teams for providing the valuable data. Y.G. and M.D.D. are supported by NSFC 
under grants 10828306 and 10933003, and by NKBRSF under grant 2011CB811402. 
The research leading to these results has received funding from the European 
Commission’s Seventh Framework Programme (FP7/2007-2013) under the grant 
agreement No. 218816 (SOTERIA project, www.soteria-space.eu). H.L. is supported 
by NSFC under grants 10873038 and 10833007, and by NKBRSF under grant 2011CB811402.


\end{CJK*}
\end{document}